# BIBI: Bayesian Inference of Breed Composition

Carlos Alberto Martínez[1*], Kshitij Khare[2†], Mauricio A. Elzo[1‡]

[1]Department of Animal Sciences
[2]Department of Statistics
University of Florida, Gainesville, FL, USA

**ABSTRACT**

The aim of this paper was to develop statistical models to estimate individual breed composition based on the previously proposed idea of regressing discrete random variables corresponding to counts of reference alleles of biallelic molecular markers located across the genome on the allele frequencies of each marker in the pure (base) breeds. Some of the existing regression-based methods do not guarantee that estimators of breed composition will lie in the appropriate parameter space and none of them account for uncertainty about allele frequencies in the pure breeds, i.e., uncertainty about the design matrix. In order to overcome these limitations, we proposed two Bayesian generalized linear models. For each individual, both models assume that the counts of the reference allele at each marker locus follow independent Binomial distributions, use the logit link, and pose a Dirichlet prior over the vector of regression coefficients (which corresponds to breed composition). This prior guarantees that point estimators of breed composition like the posterior mean pertain to the appropriate space. The difference between these models is that model termed BIBI does not account for uncertainty about the design matrix, while model termed BIBI2 accounts for such an uncertainty by assigning independent Beta priors to the entries of this matrix. We implemented these models in a dataset from the University of Florida's multibreed Angus-Brahman population. Posterior means were used as point estimators of breed composition. In addition, the ordinary least squares estimator proposed by Kuehn et al. (2011) (OLSK) was also computed. BIBI and BIBI2 estimated breed composition more accurately than OLSK, and BIBI2 had an 8.3% improvement in accuracy as compared to BIBI.

## INTRODUCTION

Due to the benefits of heterosis effects (Dickerson, 1973), crossbreeding is commonly implemented in breeding programs of many livestock and plant production systems. For example, in bovine production systems (dairy and beef) in tropical and subtropical regions of the world, many of the animals are crossbred (Elzo and Famula, 1985; Burrow and Prayaga, 2004). In animal or plants production systems using hybrid individuals, knowledge of breed composition is useful for different reasons. For instance, heterozygocity is computed using it, certain management decisions (e.g., choosing a germplasm for a given environment) are made taking it into account and it is necessary when designing crossbreeding programs (Kuehn et al., 2011; Frkonja et al., 2012; Huang et al., 2014). In addition, in several animal species the commercial value of an individual is largely determined by its breed composition and determining purity plays an important role when taking breeding decisions (Huang et al., 2014; Funkhouser et al., 2016). Furthermore, in genetic evaluation of multibreed populations, breed composition of individuals is required because it and functions of it are used as explanatory variables (Elzo and Famula, 1985). Finally, knowledge of breed composition may play an important role in certain analyses such as genome-wide association (Chiang et al., 2010; Kuehn et al. 2011).

---

[*] carlosmn@ufl.edu
[†] kdkhare@stat.ufl.edu
[‡] malezo@ufl.edu

In many populations, pedigree or ancestral breed composition data are incomplete or do not exist and the lack of any of these sources of information prevents the traditional estimation of individual breed composition. Alternatively, molecular markers can be used to estimate breed composition. This idea has been used to estimate breed composition in some species; for instance, Parker et al. (2004) used microsatellites to estimate breed composition in dogs. Hereinafter, the traditional estimator based on pedigree and ancestral breed composition records will be referred to as the pedigree-based estimator. For (not necessarily related) individuals whose parents have the same breed composition, this estimator yields the same estimate. Henceforth, groups of individuals having the same pedigree-based breed composition estimate will be denoted as breed groups. It is known that for crossbred individuals different from F1's, the actual breed composition varies from one individual to another within the same breed group due to crossing over and chromosomal assortment taking place during meiosis, and as a consequence, the pedigree-based breed composition could be far from the actual one. Thus, the use of genomic information helps in solving this problem and permits to obtain estimates of individual breed composition in the absence of pedigree or ancestral breed composition information (Chiang et al. 2010; Gorbach et al., 2010; Frkonja et al., 2012).

The availability of genotypes for thousands of SNP markers located across the whole genome permits to estimate breed composition using genomic data. Because it is very unlike that certain marker alleles exist only in a given breed (private alleles), the frequencies of the reference marker alleles in each base breed have been used (Chiang et al., 2010; Kuehn et al., 2011) to estimate breed composition. Kuehn et al. (2011) adapted the linear regression approach of Chiang et al. (2010) to estimate breed composition based on SNP genotypes using least squares. This methodology was used by Huang et al. (2014) to estimate breed composition in pig populations in the US. A limitation of this approach is that the estimators are not constrained to fall in the appropriate parameter space (a standard simplex) and consequently out-of-range estimated breed compositions can be obtained. When using the least squares method, this problem can be overcome by carrying out a linearly constrained quadratic optimization which guarantees the solution to be in the desired space. Funkhouser et al. (2016)

used this approach to estimate breed composition in the same swine populations studied by Huang et al. (2014).

A feature of this regression problem is that frequencies of reference marker alleles in base breeds are not observable and consequently they have to be estimated. Statistical methods taking uncertainty about these estimates into account have not been developed yet. Alternatively, a Bayesian estimation method permits to obtain estimates in the correct space without using constrained optimization procedures by using prior distributions having the appropriate support set. In addition, Bayesian methods provide a coherent way to take into account uncertainty about allele frequencies in base breeds.

Thus, the objective of this study was to develop statistical methods incorporating genomic data (genotypes for SNP markers located across the whole genome) and accounting for uncertainty about allele frequencies in base breeds to estimate individual breed composition using a Bayesian approach.

## MATERIALS AND METHODS

**Modelling Approaches**

The idea of estimating breed composition using a regression model whose design matrix is built using reference marker allele frequencies in base breeds (Chiang et al., 2010; Kuehn et al., 2011) is adopted here.

Let $B$ be the total number of base breeds, the objective is to estimate the fraction of each one of them in every individual. For individual $i, i = 1, 2, \ldots, n$, the fraction of breed $j$ is denoted as $\mathbb{b}_{ij}, j = 1, 2, \ldots, B$. The set of fractions of each breed for every individual $\mathbb{b}_i = (\mathbb{b}_{i1}, \mathbb{b}_{i2}, \ldots, \mathbb{b}_{iB})$ corresponds to its breed composition. For all individuals, the parameter space is the $(B-1)$-dimensional simplex

$$\Omega = \left\{ \mathbb{b} \in \mathbb{R}^B : \mathbb{b}_1, \ldots, \mathbb{b}_{B-1} \geq 0, \sum_{j=1}^{B-1} \mathbb{b}_j \leq 1, \mathbb{b}_B = 1 - \sum_{j=1}^{B-1} \mathbb{b}_j \right\} \quad (1)$$

Because of the restriction $\sum_{j=1}^{B} \mathbb{b}_{ij} = 1, \forall\, i = 1, 2, \ldots, n$, only $B - 1$ base breeds need to be considered in the analysis. The model corresponds to a Bayesian generalized linear model (GLM) in which the stochastic component is specified as follows. Let $\boldsymbol{y}_i$ be the vector containing the number of copies of the reference allele for every marker in individual $i$. Because this study focuses on diploid individuals, for individual $i$, it is assumed that $y_{ij} \sim Binomial(2, p_{ij}), 0 \leq j \leq m_i$, where $m_i$ is the number of available marker genotypes for individual $i$. Therefore, under the assumption of linkage equilibrium, the likelihood for individual $i$ corresponds to the product of $m_i$ Binomial probability mass functions. On the other hand, the canonical link (i.e., the logit link) is used; therefore, the systematic component is

$$\boldsymbol{\eta}_i = \{\eta_{ij}\}_{m_i \times 1} = \log\left(\frac{p_{ij}}{1 - p_{ij}}\right) = \boldsymbol{x}_j.\boldsymbol{\beta}_i, \quad (2)$$

where $\boldsymbol{\beta}_i$ is an unknown vector containing the regression coefficients for $B - 1$ breeds which are interpreted as the fraction of each breed in individual $i$, $X$ is a matrix containing the frequencies of the reference alleles in these $B-1$ base breeds and $\boldsymbol{x}_{j\cdot}$ is its $j^{th}$ row. Notice that $\boldsymbol{x}_{j\cdot}$ is common to all individuals genotyped for the $j^{th}$ marker locus.

Under the assumption of independence of individual genotypes, the likelihood for the complete population is

$$L(\boldsymbol{\beta};\boldsymbol{y}) = f(\boldsymbol{y}|\boldsymbol{p})$$
$$= \prod_{i=1}^{n} f(\boldsymbol{y}_i|\boldsymbol{p}_i) = \prod_{i=1}^{n} \prod_{j=1}^{m_i} f(y_{ij}|p_{ij})$$
$$\propto \prod_{i=1}^{n} \frac{\exp(\langle \boldsymbol{S}_i, \boldsymbol{\beta}_i \rangle)}{g(\boldsymbol{\beta}_i, X)} \quad (3)$$

where $\boldsymbol{y} := (\boldsymbol{y}_1', \dots, \boldsymbol{y}_n')'$, $\boldsymbol{\beta} := (\boldsymbol{\beta}_1', \dots, \boldsymbol{\beta}_n')'$, $\boldsymbol{p} := (\boldsymbol{p}_1', \dots, \boldsymbol{p}_n')'$, $\boldsymbol{p}_i := (p_{i1}, \dots, p_{im_i})'$, $1 \le i \le n$, operator $\langle \cdot, \cdot \rangle$ represents the dot product, $\boldsymbol{S}_i := \sum_{j \in M_i} y_{ij} \boldsymbol{x}_{j\cdot}$, $g(\boldsymbol{\beta}_i, X) = \prod_{j \in M_i}(1 + \exp(\langle \boldsymbol{x}_{j\cdot}, \boldsymbol{\beta}_i \rangle))^2$ and $M_i$ is the set of marker loci for which individual $i$ has been genotyped, the size of this set is $m_i$. A natural choice for the prior of $\boldsymbol{\beta}_i$ is a Dirichlet($\boldsymbol{\alpha}_i$) distribution because its support set corresponds to the appropriate parameter space; thus, $\boldsymbol{\beta}_i \overset{ind}{\sim} Dirichlet(\boldsymbol{\alpha}_i)$, $1 \le i \le n$, $\boldsymbol{\alpha}_i = (\alpha_{i1}, \dots, \alpha_{iB})$, hence:

$$\pi(\boldsymbol{\beta}_i) \propto \prod_{k=1}^{B-1} \beta_{ik}^{\alpha_{ik}-1} \left(1 - \sum_{k=1}^{B-1} \beta_{ik}\right)^{\alpha_{iB}-1}.$$

Hereinafter, this model will be referred to as BIBI.

**Remark 1.** Notice that this estimation problem has the following features. Unlike other regression models, the design matrix is not observable; therefore, it has to be estimated. The usual approach uses estimates obtained using purebred animals from a reference population (Chiang et al., 2010; Kuehn et al. 2011; Huang et al., 2014). Another feature of the problem is that for certain individuals, the true value of the parameter $\boldsymbol{\beta}_i$ is known. These are purebreds and F1 individuals. For example, suppose that the model is parameterized in terms of breeds $1, 2, \dots, B-1$, then, if individual $i$ is a purebred of breed $k$, $1 \le k \le B-1$, then, $\boldsymbol{\beta}_i = (0, \dots, 0, 1, 0, \dots, 0)'$ where scalar 1 is located in the $k^{th}$ entry of $\boldsymbol{\beta}_i$, if individual $i$ is a purebred of breed $B$ then $\boldsymbol{\beta}_i = \boldsymbol{0}_{(B-1) \times 1}$, if individual $i$ is an F1 resulting from breeds $k$ and $k'$, $1 \le k < k' \le B-1$, then the only non-null entries of $\boldsymbol{\beta}_i$ are $k$ and $k'$ each being equal to $1/2$, and so on.

Under a quadratic error loss, the Bayes estimator (i.e., the estimator minimizing the Bayesian risk) is the posterior mean (Lehmann and Casella, 1998); this is the point estimator used here. For $1 \le i \le n$, the posterior mean has the form

$$\widehat{\boldsymbol{\beta}}_i = \{\hat{\beta}_{ij}\}_{(B-1) \times 1} = E[\beta_{ij}|\boldsymbol{y}_i, X]$$
$$= \frac{\int_\Omega \beta_{ij} \prod_{k=1}^{B-1} \beta_{ik}^{\alpha_{ik}-1} (1 - \sum_{k=1}^{B-1} \beta_{ik})^{\alpha_{iB}-1} \frac{\exp(\langle \boldsymbol{S}_i, \boldsymbol{\beta}_i \rangle)}{g(\boldsymbol{\beta}_i, X)} d\boldsymbol{\beta}_i}{\int_\Omega \prod_{k=1}^{B} \beta_{ik}^{\alpha_{ik}-1} (1 - \sum_{k=1}^{B-1} \beta_{ik})^{\alpha_{iB}-1} \frac{\exp(\langle \boldsymbol{S}_i, \boldsymbol{\beta}_i \rangle)}{g(\boldsymbol{\beta}_i, X)} d\boldsymbol{\beta}_i} \quad (4)$$

Therefore, the corresponding estimator of breed composition of individual $i$ is $\widehat{\mathbb{b}}_i = \left(\widehat{\beta}_{i1}, \ldots, \widehat{\beta}_{i(B-1)}, 1 - \sum_{k=1}^{B-1} \widehat{\beta}_{ik}\right)$. By properties of the expected value (Casella and Berger, 2002), it follows that the estimated breed composition lies in $\Omega$; consequently, this method guarantees estimates in the appropriate space. Notice that this expectation can be seen as the ratio of the expectation of two functions of $\boldsymbol{\beta}_i$ taken with respect to the prior distribution, that is, with respect to a Dirichlet($\boldsymbol{\alpha}_i$). Thus,

$$\widehat{\boldsymbol{\beta}}_i = \frac{E_\pi \left[ \boldsymbol{\beta}_i \frac{\exp(\langle S_i, \boldsymbol{\beta}_i \rangle)}{g(\boldsymbol{\beta}_i, X)} \right]}{E_\pi \left[ \frac{\exp(\langle S_i, \boldsymbol{\beta}_i \rangle)}{g(\boldsymbol{\beta}_i, X)} \right]} \quad (5)$$

Consequently, if the interest is only in estimating the posterior mean, there is no need for using Markov Chain Monte Carlo (MCMC) methods because this expectation can be approximated using vanilla Monte Carlo integration (MCI). Therefore, once $N$ samples $\boldsymbol{\beta}_i^1, \ldots, \boldsymbol{\beta}_i^N$ are drawn from a Dirichlet($\boldsymbol{\alpha}_i$) distribution, the Monte Carlo approximation to $\widehat{\boldsymbol{\beta}}_i$ is

$$\widehat{\boldsymbol{\beta}}_i^{MC} = \frac{\sum_{j=1}^N \boldsymbol{\beta}_i^j \frac{\exp(\langle S_i, \boldsymbol{\beta}_i^j \rangle)}{g(\boldsymbol{\beta}_i^j, X)}}{\sum_{j=1}^N \frac{\exp(\langle S_i, \boldsymbol{\beta}_i^j \rangle)}{g(\boldsymbol{\beta}_i^j, X)}} \quad (6)$$

**Accounting for uncertainty about the design matrix.** As mentioned in Remark 1, this estimation problem exhibits the feature that the design matrix is not observable. In the previous section, a model conditioned on this matrix was proposed. The use of a Bayesian approach has the advantage of permitting to account for uncertainty about matrix $X$ in an easy way. To this end, $X$ is given a prior distribution. Recall that $x_{kj}, 1 \leq k \leq m = \max_{1 \leq i \leq n}(m_i), 1 \leq j \leq B - 1$ corresponds to the frequency of the reference allele of marker $i$ in base breed $j$, assuming independence of these random variables across breeds (i.e., independence of columns of $X$) and linkage equilibrium (i.e., independence of rows of $X$), independent Beta distributions can be posed over each entry of $X$. The other components of the model are the same; thus, the systematic component is the one shown in Equation 2, the likelihood corresponds to the one in Equation 3 and vectors $\{\boldsymbol{\beta}_i\}_{i=1}^n$, are given independent Dirichlet($\boldsymbol{\alpha}_i$) priors. Under this model, the joint posterior is:

$$\pi(\boldsymbol{\beta}, X | \boldsymbol{y}) \propto \prod_{i=1}^n \frac{\exp(\langle S_i, \boldsymbol{\beta}_i \rangle)}{g(\boldsymbol{\beta}_i, X)} \prod_{k=1}^{B-1} \beta_{ik}^{\alpha_{ik}-1}$$

$$\times \left(1 - \sum_{k=1}^{B-1} \beta_{ik}\right)^{\alpha_{iB}-1} \prod_{j=1}^m \prod_{k=1}^{B-1} x_{jk}^{a_{jk}-1} (1 - x_{jk})^{b_{jk}-1} \quad (7)$$

From Equation 7, it follows that all the full conditionals are not standard distributions and that given $\boldsymbol{\beta}$ and $\boldsymbol{y}$ the rows of $X$ are conditionally independent, but its columns are not. In this case, the dimensionality of the problem increases notably because $m$ unknown parameters are introduced in the model. Like in the previous scenario, the posterior mean can be used as a point estimator of breed composition and it can also be computed as a ratio of expectations taken with respect to the prior distribution. Let $\boldsymbol{\theta} = (\boldsymbol{\beta}', vec(X)')$ and $\Theta = \Omega \times \Gamma^{m(B-1)}$ where $vec(\cdot)$ is the vec operator and $\Gamma^{m(B-1)}$ is an $m(B-1)$-dimensional unit hypercube. Thus, the extended model accounting for uncertainty about $X$ has parameter $\boldsymbol{\theta}$ and parameter space $\Theta$ and for $1 \leq l \leq (n+m)(B-1)$:

$$\hat{\theta}_l = E[\theta_l|\boldsymbol{y}]$$

$$= \frac{\int_\Theta \theta_l L(\boldsymbol{\theta};\boldsymbol{y})\pi(\boldsymbol{\theta})\,d\boldsymbol{\theta}}{\int_\Theta L(\boldsymbol{\theta};\boldsymbol{y})\pi(\boldsymbol{\theta})\,d\boldsymbol{\theta}} = \frac{E_\pi[\theta_l L(\boldsymbol{\theta};\boldsymbol{y})]}{E_\pi[L(\boldsymbol{\theta};\boldsymbol{y})]}$$

where

$$\pi(\boldsymbol{\theta}) = \pi(\boldsymbol{\beta},X)$$

$$\propto \left(\prod_{i=1}^{n}\left(\prod_{k=1}^{B-1}\beta_{ik}^{\alpha_{ik}-1}\right)\left(1 - \sum_{k=1}^{B-1}\beta_{ik}\right)^{\alpha_{iB}-1}\right)\left(\prod_{j=1}^{m}\prod_{k=1}^{B-1}x_{jk}^{a_{jk}-1}(1-x_{jk})^{b_{jk}-1}\right)$$

and $L(\boldsymbol{\theta};\boldsymbol{y})$ is the likelihood.

In particular, the Monte Carlo approximation to the posterior mean of $\boldsymbol{\beta}_i$ based on a sample $\boldsymbol{\theta}^1,\ldots,\boldsymbol{\theta}^N$ drawn from $\pi(\boldsymbol{\theta})$ is

$$\hat{\boldsymbol{\beta}}_{2_i}^{MC} = \frac{\sum_{j=1}^{N}\boldsymbol{\beta}_i^j \frac{\exp(\langle S_i^j, \boldsymbol{\beta}_i^j\rangle)}{g(\boldsymbol{\beta}^j, X^j)}}{\sum_{j=1}^{N}\frac{\exp(\langle S_i^j, \boldsymbol{\beta}_i^j\rangle)}{g(\boldsymbol{\beta}_i^j, X^j)}} \qquad (8)$$

where $S_i^j = \sum_{k \in M_i} y_{ik} x_k^j$. This extended version of the proposed model will be referred to as BIBI2.

**Assessment of Accuracy**

Some previous studies have considered the pedigree-based breed composition estimates (Kuehn et al., 2011) or the true breed composition from simulated data (Funkhouser, 2016) to assess the adequacy of estimators of individual breed composition. When working with real data, using the second property of the estimation problem being considered in this study (see Remark 1), the accuracy of the point estimators proposed here, i.e., the posterior mean estimated using Equation 6 (when not accounting for uncertainty about $X$), or Equation 8 (when accounting for uncertainty about $X$) can be computed using the $l_1$ norm of the difference between $\boldsymbol{\beta}_i$ and $\hat{\boldsymbol{\beta}}_i$ for purebred and F1 individuals. Specifically, the following measure of accuracy was used:

$$\frac{1}{|V|}\sum_{i \in V}\|\hat{\boldsymbol{\beta}}_i - \boldsymbol{\beta}_i\|_1 \qquad (9)$$

where $V$ is the set purebred and F1 individuals, $|V|$ its cardinality (number of elements) and operator $\|\cdot\|_1$ represents the $l_1$ norm.

**Implementation in a Multibreed Bovine Population**

Implementation with real data was carried out using a multibreed beef cattle population from the University of Florida's Beef Research Unit (BRU). This multibreed population was created in 1989 and it is composed of two breeds: Angus and Brahman. Individuals are mated according to a diallel design; therefore, this population contains individuals with breed composition ranging from purebred Angus to purebred Brahman. Details on this multibreed population can be found in Elzo and Wakeman (1998). Data came from a subset of 120 individuals with pedigree-based estimated Angus fraction ranging from 0 to 1 that were genotyped for SNP markers included in the GeneSeek Genomic Profiler F-250 (Neogen Corporation, Lansing, MI, USA). Only those markers in common with the Illumina BovineSNP50 array (Illumina Inc., San Diego, CA, USA) were considered. In order to construct the design matrix for model BIBI, and to define the prior for $X$ in model BIBI2, the reference allele frequencies in Angus and Brahman reported by Kuehn et al. (2011) were used. These values were the entries of the design matrix when fitting BIBI whereas when fitting BIBI2 the hyperparameters were chosen in such a way that prior means matched them. Markers located in autosomes that had a minor allele frequency larger than 0.05 were considered. Then, marker loci for which all individuals were heterozygous were removed. After this editing, a total of 9906 markers were left. In addition to models BIBI and BIBI2, the model proposed by Kuehn et al. (2011) was fitted to this data. Accuracy of estimators was assessed using Equation 9 in a set of 60 individuals known to be purebreds (Angus and Brahman) and F1. Henceforth this set will be referred to as MAB-V. In addition to the average in Equation 9, other descriptive statistics of the norm $\|\widehat{\boldsymbol{\beta}}_i - \boldsymbol{\beta}_i\|_1$ were also computed. Finally, Pearson correlation coefficients among the different estimators of Angus fraction were computed in the complete set of 120 individuals and in the subset of 60 non-F1 crossbred individuals.

## RESULTS

According to the expression presented in Equation 9, from the three models considered here, BIBI2 had the best performance and the estimator proposed by Kuehn et al. (2011) had the worst. Moreover, estimates from BIBI2 were less variable across individuals than those obtained from the other two models (Table 1).

Table 1. Descriptive statistics of the $l_1$ norm of the difference between estimated and true breed composition in the MAB-V set.

| Statistic | BIBI | BIBI2 | KOLS[*] |
| --- | --- | --- | --- |
| Mean | 0.026 | 0.024 | 0.200 |
| Standard deviation | 0.003 | 0.001 | 0.052 |
| Minimum | 0.011 | 0.021 | 0.067 |
| Maximum | 0.028 | 0.026 | 0.266 |

[*]Ordinary least squares estimator based on the model proposed by Kuehn et al. (2011)

On the other hand, correlations between estimated Angus fractions from the four estimators considered here (those from the three regression models mentioned above and the pedigree-based estimator) were high in both the complete set of 120 individuals and the subset of 60 non-F1 crossbred individuals (Table 2).

Table 2. Pearson correlation coefficients among different kinds of Angus fraction estimates in the complete set and in the subset of non-F1 crossbred individuals[*]

|                | BIBI | BIBI2 | KOLS | Pedigree-based |
|----------------|------|-------|------|----------------|
| BIBI           | 1    | 0.94  | 0.92 | 0.94           |
| BIBI2          | 0.95 | 1     | 0.92 | 0.94           |
| KOLS           | 0.92 | 0.92  | 1    | 0.92           |
| Pedigree-based | 0.95 | 0.95  | 0.92 | 1              |

[*]Upper off-diagonal elements correspond to correlations in the subset of non-F1 crossbred individuals while lower off-diagonal elements correspond to correlations in the complete set.

The smallest value was 0.92, it corresponded to the correlation between the estimates from the method proposed by Kuehn et al. (2011) and the pedigree-based estimates in the two sets, and the largest was 0.95, corresponding to the correlations between BIBI and the pedigree-based estimates and between BIBI and BIBI2 estimates in the complete set (Table 2).

## DISCUSSION

In this study, Bayesian GLM's to estimate breed composition in diploid species using biallelic molecular markers were developed. Specifically, these models correspond to Bayesian logistic regressions, assigning a Dirichlet prior to the vector of regression coefficients. This prior was chosen because its support set matches the parameter space of this estimation problem which is the simplex presented in Equation 1. In a first stage, the model was formulated conditional on the design matrix $X$ (BIBI), which is formed using marker allele frequencies in base breeds. However, due to the fact that these frequencies are unknown, this estimation problem has as special feature the fact that the design matrix is not observable; therefore, it has to be estimated. Consequently, this model was expanded to account for uncertainty about the design matrix (BIBI2); this expanded version treats the design matrix as an unknown parameter and independent Beta priors are assigned to each one of its entries. Thus, this model takes into account the fact that there exists uncertainty about the design matrix and incorporates it in the estimation process. However, the model conditional on the design matrix is simpler because for each individual it only has B-1 parameters where B is the number of base breeds. Consequently, both versions of the model were considered. The point estimator used here was the posterior mean, which according to decision theory, minimizes the Bayesian risk under the squared error loss (Lehmann and Casella, 1998). In both models, the posterior and full conditional distributions are unknown.

In a regression model, when the explanatory variables are recorded with errors, it could induce bias in the parameter estimates; therefore, there exist approaches to take the measurement error in regression variables into account (Lehmann and Casella, 1998; Rawlings et al., 1998). In a similar spirit, in the problem addressed here, the regression variables are actually estimates, thus, they may differ from the true values. Hence, the model accounting for this fact is theoretically more rigorous and it could be expected to yield more accurate estimators of breed composition. A small increment in accuracy (as measured by the expression in Equation 9) was observed in this study (Table 1). Relative to the accuracy of method BIBI, it corresponds to an improvement of 8.3%. At this point, it needs to be mentioned that beyond the potential gains in accuracy and their magnitude, model BIBI2 features the following practical advantages. Firstly, when a reference population to estimate allele frequencies in base breeds is not available, BIBI2 still permits using data from multibreed populations to estimate individual breed composition. Secondly, if it is of interest, estimates of the entries of the design matrix, namely, allele frequencies in base populations can be computed because these are model parameters in BIBI2. Consequently, BIBI2 can be regarded as a more versatile and theoretically rigorous model.

Bayesian estimation of breed composition has another appealing feature. In certain cases there is interest in making probabilistic statements about breed composition; for example, to answer the question: what is the probability that the fraction of breed $k$ in individual $i$ is larger than $t$? Models like BIBI and BIBI2 permit to easily answer this kind of questions by using the posterior distribution of $\boldsymbol{\beta}$. Specifically, if $\Psi$ is a proper subset of $\Omega$, then

$$P(\boldsymbol{\beta} \in \Psi | \boldsymbol{y}) = \int_{\Psi} \pi(\boldsymbol{\beta}|\boldsymbol{y}) d\boldsymbol{\beta} \qquad (10)$$

Notice that for model BIBI2 obtaining $\pi(\boldsymbol{\beta}|\boldsymbol{y})$ requires integration of the joint posterior density of $\boldsymbol{\beta}$ and $X$ with respect to $X$ and that for models BIBI and BIBI2, the integral in Equation 10 has to be calculated numerically. To this end, a Markov Chain Monte Carlo method can be used to compute an approximation to $P(\boldsymbol{\beta} \in \Psi | \boldsymbol{y})$. In addition, interval estimates of breed composition can be obtained by computing credible sets.

Regarding evaluation of accuracy of the proposed estimators, using the fact that in this estimation problem the true value of $\boldsymbol{\beta}_i$ is known for purebred and F1 individuals, it was proposed to use this set to assess the accuracy of a given estimator of breed composition. An approach in the same spirit, but considering only purebred individuals from some breed of interest has been used in previous studies (Frkonja et al., 2012; Huang et al., 2014, Funkhouser et al., 2016). In particular, the $l_1$ norm was used here, but other norms such as the $l_2$ norm could be used as well, the key is taking advantage of the aforementioned particular feature of the problem being addressed. Notice that for purebred and F1 individuals the pedigree-based estimator is 100% accurate; thus, Equation 9 is useful to compare marker-based estimators because the pedigree-based estimator will always have a value of zero. A drawback of this measure is that it does not consider the whole parameter space; it only assesses the performance of different estimators at certain points of this space. Due to the biological reasons explained in the introduction, marker-based estimators could be expected to outperform the pedigree-based estimator in non-F1 crossbred individuals because they consider individual variation in breed composition; recall that the pedigree-based estimator takes the same value for all the offspring of a given pair of parents whereas a marker-based estimator does not. Kuehn et al. (2011) used the pedigree-based estimates of breed composition of crossbred individuals to assess the performance of their regression-based estimator. However, as mentioned before, due to crossing over and chromosome assortment, the pedigree-based estimates of breed composition could be far from the true values and consequently this approach may not be entirely appropriate to assess accuracy. Therefore, simulation is a valuable tool to assess the accuracy of different estimators of breed composition. Funkhouser et al. (2016) carried out a simulation study to assess the performance of their constrained regression method and they found correlations between estimated and true breed compositions of 0.97.

High correlations between estimates of Angus fraction from the four methods considered here were found, in particular, estimates from all regression-based methods had high correlations with the pedigree-based estimates (Table 2). This result is in agreement with results reported by Kuehn et al. (2011) and Frkonja et al. (2012). For the set of individuals analyzed in this study, although correlations between estimates of Angus fraction obtained from the Bayesian estimators and the least squares estimator proposed by Kuehn et al. (2001) were high, the accuracy of the Bayesian estimators was considerably higher. However, notice that the accuracy of the least squares estimator was reasonably high (Table 1).

As discussed previously, marker-based estimators of individual breed composition are a useful and reliable tool to infer breed composition when pedigree and/or ancestral breed composition records are not available. However, the benefits of marker-based estimators go beyond this scenario because having a better knowledge of individual breed composition has several potential applications even when pedigree and ancestral breed composition are known. The following are some examples of problems in which having marker-based estimates of breed composition could have a positive impact as compared to pedigree-based estimates. Selection of candidates for crossbreeding programs, creation of the so-called "synthetic breeds" or creation of genetic lines, and estimation of heterosis effects (Dickerson, 1973), average breed additive effects and multibreed genetic values (Elzo and Famula, 1985). As to the latter topic, the fact that for non-F1 crossbred individuals pedigree-based breed composition estimates could be far apart from the actual composition poses a question regarding the impact of the use of marker-based estimates on the accuracy of predicted multibreed genetic values as defined in Elzo and Famula (1985). Breed composition and functions of it are used as explanatory variables in the linear models used to predict genetic values and to estimated genetic parameters in multibreed populations (Elzo and Famula, 1985, Elzo 1994, Elzo, 1996; Cardoso and Tempelman, 2004); therefore, it could be of interest to compare the performance of a multibreed model considering marker-based estimated breed compositions with the standard model that uses pedigree-based estimates. An example of a successful application of marker-based estimation of individual breed composition is the adoption of the methods developed by Funkhouser et al. (2016) by the US National Swine Registry (NSR) to screen purebred Yorkshire pigs.

A refinement of models BIBI and BIBI2 is to consider kinship. It can be accomplished by removing the assumption of conditional independence of vectors $\boldsymbol{y}_1, \dots, \boldsymbol{y}_n$ given $X$ and $\boldsymbol{\beta}$ and modifying the joint probability mass function of these vectors, i.e., the likelihood function. When pedigree information is available, a likelihood based on genetic relationships can be derived using the pedigree by following a derivation similar to the one presented in Martínez et al. (2017).

An alternative approach to estimate breed composition is to use distance-based or model-based clustering methods that have been designed to infer population structure.

Examples of software implementing this sort of methods are STRUCTURE (Pritchard et al., 2000), MENDEL (Lange et al., 2001) and EigenStrat (Price et al., 2006). Most of them can perform what is known as a soft clustering, that is, instead of assigning individuals to clusters (populations), they compute the probability that each individual pertains to each cluster (Hastie et al., 2009). Because the focus of this study was on the regression approach, these alternative methods were not implemented. Moreover, previous studies have reported high degree of agreement among the estimates from some of these methods and the regression approach (Chiang et al., 2010; Kuehn et al., 2011). Notwithstanding, it is worth mentioning the following drawback of some of these clustering methods. They are designed to infer a "cryptic" population structure, which is useful when the underlying factors inducing genetic heterogeneity (structure) in the population are not easily detected. In fact, they can infer the number of subpopulations (i.e., the number of clusters); however, when using them to estimate breed composition, the number of clusters is set equal to the number of breeds. Thus, for each individual what is estimated is the probability of pertaining to each one of these clusters; consequently, breed fractions are not estimated directly and clusters have to be assigned to breeds, e.g., cluster one corresponds to breed B and cluster 2 corresponds to breed A. Usually, this assignment is performed by using purebred individuals or individuals known to have a high frequency of certain breed. For two or three breeds, this is not too difficult, but when the number of breeds increases and pedigree-based estimates and purebred individuals of certain breeds are not available, assigning clusters to breeds could be difficult and this may introduce additional bias. For example, consider the problem of estimating breed composition in a crossbred population composed of eight breeds where purebred individuals and pedigree are not available.

## FINAL REMARKS

The use of genomic data permits to estimate breed composition when pedigree or ancestral breed composition information are missing, but advantages of this kind of estimators go beyond this. As an example, consider the potential benefits of the use of marker-based estimates of breed composition in different genetics problems that were discussed here. Therefore, this approach could be useful even when pedigree and ancestral breed composition records are available. Hence, the potential applications of these estimators open a path for further research. The regression models developed here permit to easily obtain point and interval estimates of breed composition and to make useful probabilistic statements about it. Also, unlike other regression models used to estimate breed composition, model BIBI2 accounts for uncertainty about allele frequencies in base breeds and permits to estimate breed composition in the absence of reference populations to estimate them.

## ACKNOWLEDGEMENTS

Funds to obtain the data used in this study came from the following research projects: Hatch Project FLA-ANS- 005150, Florida Beef Council Project 00098682, and Florida Cattle

Enhancement Fund Project 00125199. Authors thank Dr. Raluca Mateescu from the Department of Animal Sciences of University of Florida for providing genotypic data and Dr. Juan Pedro Steibel from the Department of Animal Science and the Department of Fisheries and Wildlife of Michigan State University for useful discussions. C.A. Martínez thanks Fulbright Colombia and "Departamento Adiministrativo de Ciencia, Tecnología e Innovación" COLCIENCIAS for supporting his PhD and master's programs at the University of Florida through a scholarship, and Bibiana Coy for her love, support and constant encouragement.

## REFERENCES

Burrow, H.M., Prayaga, K.C., 2004. Correlate responses in productive and adaptative traits and temperament following selection for growth and heat resistance in tropical beef cattle. Livest. Prod. Sci. 86, 143-161.

Cardoso, F.F., Tempelman, R.J., 2004. Hierarchical Bayes multiple-breed inference with an application to genetic evaluation of a Nelore-Hereford population. J. Anim. Sci. 82, 1589-1601.

Chiang, C.W.K., Gajdos, Z.K.Z., Korn, J.M., Kuruvilla, F.G., Butler, J.L., Hackett, R., Guiducci, C., Nguyen, T.T., Wilks, R., Forrester, T., Haiman, C.A., Henderson, K.D., Le Marchand, L., Henderson, B.E., Palmert, M.R., McKenzie, C.A., Lyon, H.N., Cooper, R.S., Zhu, X., & Hirschhorn, J.N., 2010. Rapid Assessment of Genetic Ancestry in Populations of Unknown Origin by Genome-Wide Genotyping of Pooled Samples. Plos Genet. 6(3), e1000866.

Dickerson, G.E., 1973. Inbreeding and heterosis in animals. Proceedings of the Animal Breeding and Genetics Symposium in Honor of Dr. J.L. Lush, 54-77.

Elzo M.A., Famula, T.R., 1985. Multibreed sire evaluation within a country. J. Anim. Sci. 60, 942-952.

Elzo, M.A., 1994. Restricted maximum likelihood estimation of additive and nonadditive genetic variances and covariances in multibreed populations. J. Anim. Sci. 72, 3055-3065.

Elzo, M.A., 1996. Unconstrained procedures for the estimation of positive definite covariance matrices using restricted maximum likelihood in multibreed populations. J. Anim. Sci. 74, 317-328.

Elzo, M.A., Wakeman, D.L., 1998. Covariance components and prediction for additive and nonadditive preweaning growth genetic effects in an Angus-Brahman multibreed herd. J. Anim. Sci. 76, 1290-1302.

Frkonja, A., Gredler, B., Schnyder, U., Curik, I., Solkner, J., 2012. Prediction of breed composition in an admixed cattle population. Anim. Genet. 43, 696-703.

Gorbach, D.M., Makgahlela, M.L., Reecy, J.M., Kemp, S.J., Baltenweck, I., Ouma, R., Mwai, O., Marshall, K., Murdoch, B., Moore, S., Rothschild, M.F., 2010. Use of SNP genotyping to determine pedigree and breed composition of dairy cattle in Kenya. J. Anim. Breed. Genet. 127, 348-351.

Hastie, T., Tibshirani, R., Friedman, J., 2008. The elements of statistical learning. Data mining, inference, and prediction 2nd ed. Springer, New York, NY, USA.

Huang, Y., Bates, R.O., Ernst, C.W., Fix, J.S., Steibel, J.P., 2014. Estimation of U.S. Yorkshire breed composition using genomic data. J. Anim. Sci. 92, 1395-1404.

Kuehn, L.A., Keele, J.W., Bennett, G.L., McDaneld, T.G., Smith, T.P.L., Snelling, W.M., Sonstegard, T.S., Thallman, R.M., 2011. Predicting breed composition using breed frequencies of 50,000 markers from the US Meat Animal Research Center 2,000 Bull Project. J. Anim. Sci. 89, 1742-1750.

Lange, K., Cantor, R., Horvath, S., Perola, M., Sabatti, C., Sinsheimer, J., Sobel, E., 2001. Mendel version 4.0: A complete package for the exact genetic analysis of discrete traits in pedigree and population datasets. Am. J. Hum. Genet. 69(Suppl), 504.

Martínez, C.A., Khare, K., Banerjee, A., Elzo, M.A., 2016. Joint genome-wide prediction in several populations accounting for randomness of genotypes: A hierarchical Bayes approach. I: Multivariate Gaussian priors for marker effects and derivation of the joint probability mass function of genotypes. J. Theor. Biol. 417, 8-19.

Parker, H.G., Kim, L.V., Sutter, N.B., Carlson S., Lorentzen T.D., Malek, T.B., Johnson, G.S., DeFrance, H.B., Ostrander, E.A., Kruglyak, L., 2004. Genetic structure if the purebred domestic dog. Science 304, 1160-1164.

Price, A.L., Patterson, N.J., Plenge, R.M., Weinblatt, M.E., Shadick, N.A., Reich, D., 2006. Principal components analysis correct for stratification in genome-wide association studies. Nat. Genet. 38, 904-909.

Pritchard, J.K., Stephens, M., Donnelly, P., 2000. Inference of population structure using multilocus genotype data. Genetics 155, 945-959.

Rawlings, J.O., Pantula, S.G., Dickey, D.A., 1998. Applied regression analysis. A research tool 2nd ed. Springer, New York, NY, USA.

# Appendix
# Toy example of computations of point estimators of breed composition using Monte Carlo Integration

Suppose you have a multibreed population composed of two breeds. Individual number 3 is genotyped for 10 biallelic molecular markers and the following are the counts of the reference allele at each marker locus.

| Marker locus | Reference allele count |
|---|---|
| 1 | 1 |
| 2 | 2 |
| 3 | 0 |
| 4 | 1 |
| 5 | 1 |
| 6 | 2 |
| 7 | 0 |
| 8 | 1 |
| 9 | 2 |
| 10 | 0 |

Thus, $y_3 = (1\ 2\ 0\ 1\ 1\ 2\ 0\ 1\ 2\ 0)'$. Also, assume that models BIBI and BIBI2 are parameterized in terms of breed A, and that the following are the reference allele frequencies previously estimated in a reference population of purebred A individuals: 0.40, 0.11, 0.23, 0.30, 0.80, 0.97, 0.78, 0.56, 0.12, 0.01 for marker loci 1, 2,…,10 respectively. Consequently, when fitting model BIBI the design matrix is just a column vector, that is,

$$X = (0.40\ 0.11\ 0.23\ 0.30\ 0.80\ 0.97\ 0.78\ 0.56\ 0.12\ 0.01)'$$

In this toy example, these imaginary estimates of allele frequencies are the counterpart of those reported by Kuehn et al. (2011) in the manuscript.

Suppose that estimator shown in Equation 5 (which comes from model BIBI) is based on 10 samples from a Beta (1,1) distribution. Recall that a Beta(1,1,) distribution is equivalent to a continuous Uniform (0,1) distribution. Of course, the small number of samples obeys to the fact that this is just an illustration. The following is the sample:

0.6357237, 0.7692941, 0.3054812, 0.6153925, 0.2540863, 0.3890438, 0.9892577, 0.4186421, 0.5969372, 0.9848212 which corresponds to $\beta_3^1, \beta_3^2, \ldots, \beta_3^{10}$, respectively.

Then the estimator obtained from model BIBI shown in Equation 5 is as follows:

$$\hat{\beta}_3^{MC} = \frac{\sum_{j=1}^{10} \beta_3^j \frac{\exp(S_3 \beta_3^j)}{g(\beta_3^j, X)}}{\sum_{j=1}^{10} \frac{\exp(S_3 \beta_3^j)}{g(\beta_3^j, X)}}$$

Notice that in this case $S_3$ and $\beta_3^1, \beta_3^2, \ldots, \beta_3^{10}$ are real numbers, therefore the $\langle S_3, \beta_3^j \rangle = S_3 \beta_3^j$ for all $j$.

The numerator is

$$0.6357237 \frac{\exp(0.6357237 S_3)}{g(0.6357237, X)} + 0.7692941 \frac{\exp(0.7692941 S_3)}{g(0.7692941, X)} + \cdots$$
$$+ 0.9848212 \frac{\exp(0.9848212 S_3)}{g(0.9848212, X)}$$

whereas the denominator is

$$\frac{\exp(0.6357237 S_3)}{g(0.6357237, X)} + \frac{\exp(0.7692941 S_3)}{g(0.7692941, X)} + \cdots + \frac{\exp(0.9848212 S_3)}{g(0.9848212, X)}$$

where

$$S_3 = \sum_{j=1}^{10} y_{3j} x_j = 1 \times 0.40 + 2 \times 0.11 + \cdots + 0 \times 0.01$$

$$g(0.6357237, X) = \prod_{j=1}^{10} \left(1 + \exp(0.6357237 x_j)\right)^2$$
$$= (1 + \exp(0.6357237 \times 0.40))^2 (1 + \exp(0.6357237 \times 0.11))^2 \cdots (1 + \exp(0.6357237 \times 0.01))^2$$

$$g(0.7692941, X) = \prod_{j=1}^{10} \left(1 + \exp(0.7692941 x_j)\right)^2$$
$$= (1 + \exp(0.7692941 \times 0.40))^2 (1 + \exp(0.7692941 \times 0.11))^2 \cdots (1 + \exp(0.7692941 \times 0.01))^2$$

$$\vdots$$

$$g(0.9848212, X) = \prod_{j=1}^{10} \left(1 + \exp(0.9848212 x_j)\right)^2$$
$$= (1 + \exp(0.9848212 \times 0.40))^2 (1 + \exp(0.9848212 \times 0.11))^2 \cdots (1 + \exp(0.9848212 \times 0.01))^2$$

After carrying out all the computations shown above the result is $\hat{\beta}_i^{MC} = 0.0596$; therefore, the estimated fraction of breed B is $1 - 0.0596 = 0.9404$.

Now, let's focus on the estimator from model BIBI2 shown in Equation 7.

$$\hat{\beta}_{2_3}^{MC} = \frac{\sum_{j=1}^{10} \beta_3^j \frac{\exp(S_3^j \beta_3^j)}{g(\beta_3^j, X^j)}}{\sum_{j=1}^{10} \frac{\exp(S_3^j \beta_3^j)}{g(\beta_3^j, X^j)}}$$

The same sample $\beta_3^1, \beta_3^2, \ldots, \beta_3^{10}$ is used, but this time we need to obtain samples of the design matrix. To this end, we use matrix $X$ presented above to define the hyper-parameters. This prior corresponds to the product of ten Beta distributions, one per marker. For each element of X, the hyper-parameters of the corresponding prior are chosen such that the mean of this distribution matches the allele frequencies estimated in breed A and a desired variance or a coefficient of variation. Suppose that for the ten markers, there exists high certainty about the estimates of allele frequencies in base breeds and consequently a coefficient of variation (CV) of 0.03 is set for all marker loci. Thus, for each marker locus, using the first two moments of a Beta($a_i, b_i$) distribution, $i = 1,2,\ldots,10$, the following equations have to be solved:

$$\mu_i = \frac{a_i}{a_i + b_i}$$
$$\sigma_i^2 = \frac{a_i b_i}{(a_i + b_i)^2 (a_i + b_i + 1)}$$

where $\mu_i$ and $\sigma_i^2$ are the mean and variance of the prior distribution of the reference allele frequency of the $i^{th}$ maker locus ($i = 1,2,\ldots,10$) in breed A. Consequently, $\mu_i$ corresponds to $X_i$. On the other hand, $\sigma_i^2$ is easily determined from $\mu_i$ and $CV_i$ as $\sigma_i^2 = (CV_i \mu_i)^2$, since $CV_i = 0.03$ for all $i$ and $\mu_i = X_i$, it follows that $\sigma_i^2 = (0.03 X_i)^2$.
From the two equations presented above it follows that:

$$a_i = \mu_i \left( \frac{\mu_i(1 - \mu_i)}{\sigma_i^2} - 1 \right)$$
$$b_i = \alpha_i \left( \frac{1}{\mu_i} - 1 \right)$$

For example, for the first marker locus $\mu_1 = X_1 = 0.40$, $\sigma_1^2 = (0.03 \times 0.40)^2 = 0.000144$, $a_1 = 666.2667, b_1 = 999.4$. As a check, notice that

$$\frac{666.2667}{666.2667 + 999.4} = 0.4$$

$$\frac{666.2667 \times 999.4}{(666.2667 + 999.4)^2 (666.2667 + 999.4 + 1)} = 0.000144$$

The following are the hyper-parameters of the prior distributions of the reference allele frequencies in breed A for the ten marker loci:

| Marker locus | $a_i$ | $b_i$ |
|---|---|---|
| 1 | 666.2667 | 999.4 |
| 2 | 988.7789 | 8000.12 |
| 3 | 855.3256 | 2863.481 |
| 4 | 777.4778 | 1814.115 |
| 5 | 221.4222 | 55.35556 |
| 6 | 32.36333 | 1.000928 |
| 7 | 243.6644 | 68.72587 |
| 8 | 488.3289 | 383.687 |
| 9 | 977.6578 | 7169.49 |
| 10 | 1099.99 | 108899 |

Now, ten samples are drawn from each one of the prior distributions of reference allele frequencies resulting in the following array

| Marker locus | Sample | | | | | | | | | |
|---|---|---|---|---|---|---|---|---|---|---|
| | 1 | 2 | 3 | 4 | 5 | 6 | 7 | 8 | 9 | 10 |
| 1 | 0.395332 | 0.394856 | 0.425437 | 0.380636 | 0.409263 | 0.388591 | 0.391048 | 0.392275 | 0.371358 | 0.386719 |
| 2 | 0.112126 | 0.108584 | 0.116923 | 0.112506 | 0.111977 | 0.104564 | 0.116228 | 0.108135 | 0.110356 | 0.112664 |
| 3 | 0.233824 | 0.224962 | 0.226910 | 0.236674 | 0.234249 | 0.227868 | 0.235572 | 0.223145 | 0.236582 | 0.224286 |
| 4 | 0.300981 | 0.286014 | 0.283136 | 0.310659 | 0.299265 | 0.300990 | 0.319340 | 0.314176 | 0.301910 | 0.300894 |
| 5 | 0.839877 | 0.817418 | 0.784380 | 0.771902 | 0.768407 | 0.826425 | 0.778987 | 0.793831 | 0.823895 | 0.815640 |
| 6 | 0.938834 | 0.983531 | 0.981287 | 0.994615 | 0.989447 | 0.989435 | 0.890145 | 0.958794 | 0.973400 | 0.887737 |
| 7 | 0.776349 | 0.813679 | 0.778738 | 0.758412 | 0.766998 | 0.797483 | 0.750849 | 0.795898 | 0.770344 | 0.784363 |
| 8 | 0.579811 | 0.573914 | 0.555975 | 0.535200 | 0.560810 | 0.552444 | 0.544482 | 0.522569 | 0.582289 | 0.547437 |
| 9 | 0.125026 | 0.114989 | 0.124421 | 0.118356 | 0.128008 | 0.122250 | 0.124481 | 0.117951 | 0.114007 | 0.119690 |
| 10 | 0.009804 | 0.010026 | 0.009571 | 0.009705 | 0.010310 | 0.009945 | 0.009847 | 0.009360 | 0.009850 | 0.010233 |

Hence, the $j^{th}$ sample used to compute $\hat{\beta}_{2_3}^{MC}$ corresponds to the $j^{th}$ column of the table shown above (denoted as $X^j$) and $\beta_3^j, j = 1,2,...,10$. Therefore, the numerator of $\hat{\beta}_{2_3}^{MC}$ is

$$0.6357237\frac{\exp(0.6357237 S_3^1)}{g(0.6357237, X^1)} + 0.7692941\frac{\exp(0.7692941 S_3^2)}{g(0.7692941, X^2)} + \cdots$$
$$+ 0.9848212\frac{\exp(0.9848212 S_3^{10})}{g(0.9848212, X^{10})}$$

and the denominator is

$$\frac{\exp(0.6357237 S_3^1)}{g(0.6357237, X^1)} + \frac{\exp(0.7692941 S_3^2)}{g(0.7692941, X^2)} + \cdots + \frac{\exp(0.9848212 S_3^{10})}{g(0.9848212, X^{10})}$$

where

$$S_3^1 = \sum_{j=1}^{10} y_{3j} x_j^1 = 1 \times 0.395332 + 2 \times 0.112126 + \cdots + 0 \times 0.009804$$

$$S_3^2 = \sum_{j=1}^{10} y_{3j} x_j^2 = 1 \times 0.394856 + 2 \times 0.108584 + \cdots + 0 \times 0.010026$$

$$\vdots$$

$$S_3^{10} = \sum_{j=1}^{10} y_{3j} x_j^{10} = 1 \times 0.386719 + 2 \times 0.112664 + \cdots + 0 \times 0.010233$$

$$g(0.6357237, X^1) = \prod_{j=1}^{10} \left(1 + \exp(0.6357237 x_j^1)\right)^2$$
$$= (1 + \exp(0.6357237 \times 0.395332))^2 (1 + \exp(0.6357237 \times 0.112126))^2 \cdots (1 + \exp(0.6357237 \times 0.009804))^2$$

$$g(0.7692941, X^2) = \prod_{j=1}^{10} \left(1 + \exp(0.7692941 x_j^2)\right)^2$$
$$= (1 + \exp(0.7692941 \times 0.394856))^2 (1 + \exp(0.7692941 \times 0.108584))^2 \cdots (1 + \exp(0.7692941 \times 0.010026))^2$$

$$\vdots$$

$$g(0.9848212, X^{10}) = \prod_{j=1}^{10} \left(1 + \exp(0.9848212 x_j^{10})\right)^2$$
$$= (1 + \exp(0.9848212 \times 0.386719))^2 (1 + \exp(0.9848212 \times 0.112664))^2 \cdots (1 + \exp(0.9848212 \times 0.010233))^2$$

After carrying out these computations the estimated fraction of breed A is 0.0595 and therefore the estimated fraction of breed B is 0.9405.